\documentclass[aps,prd,groupedaddress, twocolumn,  nofootinbib]{revtex4-1}
\usepackage{graphicx, amsmath, bbm}

\begin{document}

\title{Investigation of the Spinfoam Path integral with Quantum Cuboid Intertwiners}

\author{Benjamin Bahr}
\email[]{benjamin.bahr@desy.de}

\author{Sebastian Steinhaus}
\email[]{sebastian.steinhaus@desy.de}
\affiliation{II. Institute for Theoretical Physics\\University of Hamburg\\ Luruper Chaussee 149\\22761 Hamburg\\Germany}

\date{\today}

\begin{abstract}
In this work, we investigate the 4d path integral for Euclidean quantum gravity on a hypercubic lattice, as given by the Spin Foam model by Engle, Pereira, Rovelli, Livine, Freidel and Krasnov (EPRL-FK). To tackle the problem, we restrict to a set of quantum geometries that reflects the large amount of lattice symmetries. In particular, the sum over intertwiners is restricted to quantum cuboids, i.e.~coherent intertwiners which describe a cuboidal geometry in the large-$j$ limit.

Using asymptotic expressions for the vertex amplitude, we find several interesting properties of the state sum. First of all, the value of coupling constants in the amplitude functions determines whether geometric or non-geometric configurations dominate the path integral. Secondly, there is a critical value of the coupling constant $\alpha$, which separates two phases. In both phases, the diffeomorphism symmetry appears to be broken. In one, the dominant contribution comes from highly irregular, in the other from highly regular configurations, both describing flat Euclidean space with small quantum fluctuations around them, viewed in different coordinate systems. On the critical point diffeomorphism symmetry is nearly restored, however.

Thirdly, we use the state sum to compute the physical norm of kinematical states, i.e.~their norm in the physical Hilbert space. We find that states which describe boundary geometry with high torsion have exponentially suppressed physical norm. We argue that this allows one to exclude them from the state sum in calculations.
\end{abstract}


\maketitle

\section{Motivation}

The spin foam approach has been developed to give a rigorous meaning to the path integral for quantum gravity (see \cite{Perez:2012wv} for a review). Its central idea rests on the observation  that the first order formalism of GR can be rewritten as a certain constrained topological theory \cite{Plebanski:1977zz, Horowitz:1989ng}, dubbed ``BF theory''. A quantization choice for these constraints is what specifies the spin foam model, and in recent years there have been several proposals \cite{Barrett:1997gw, Engle:2007wy, Freidel:2007py, Baratin:2011hp}. A popular choice has emerged in the so-called EPRL-FK model \cite{Engle:2007wy, Pereira:2007nh, Engle:2008ev}, which possesses quite useful properties. In particular, the resulting amplitude has an asymptotic expression for large quantum numbers which reproduces the Regge action \cite{Barrett:2009gg, Barrett:2009mw, Conrady:2008mk}. Furthermore, the resulting path integral for the EPRL-FK model naturally has boundary states which resemble the spin network states from canonical loop quantum gravity \cite{cbook,  thomasbook, Rovelli:2010km}, which is why it has been coined ``covariant loop quantum gravity'' (\cite{Rovelli:2010wq}, see also \cite{Alesci:2011ia, Thiemann:2013lka}).

There are several open questions, however. While there are numerous results available which elucidate the property of a single vertex amplitude, very little is known about the  behaviour of the whole path integral.\footnote{There are some results on the asymptotic expression for more than one vertex \cite{Bonzom:2009hw, Hellmann:2013gva}, as well as self-energy calculations \cite{Riello:2013bzw}.} In particular, it is the realm of many building blocks which is of utmost importance if one wants to understand the continuum limit of the theory. This is closely connected to the question of renormalizability of the model, as well as the challenge of restoring the broken diffeomorphism invariance \cite{Dittrich:2008pw, Bahr:2009ku, Bahr:2009qc}, both of which are unsolved questions up to this point.

The non-perturbative and background-independent nature of loop quantum gravity suggests that numerical methods from lattice gauge theory could be very useful tools for investigations, as well as extracting physical predictions from the theory. This assumption is supported by the fact that spin foam models are by construction generalized lattice gauge theories \cite{Bahr:2012qj}.\footnote{A complementary but closely connected road towards the understanding of the continuum limit comes from the group field theory / tensor field theory  approach \cite{Oriti:2007qd, Rivasseau:2011hm}, which is using much more the particle physics understanding of the spin foam amplitudes.}

There has been considerable progress in recent years on understanding the continuum limit of so-called spin net models, which are analogues of spin foam models \cite{Dittrich:2011zh}. Using numerical algorithms in tensor network renormalization \cite{levin, guwen}, many results about the phase structure of these models, as well as their continuum limit properties, can be derived, in particular in the realm of finite and quantum groups \cite{Dittrich:2013bza, Dittrich:2013voa}. The methods can also applied to lattice gauge theories \cite{Dittrich:2014mxa}.

This, together with the fact that many results in lattice gauge theory are accessed numerically, in particular high-precision predictions of physical quantities, suggests that the numerical investigation of the full spin foam path integral will be a crucial step on the road to a predictive quantum gravity theory, as well as in understanding its continuum limit and renormalization group flow \cite{Bahr:2014qza, Dittrich:2014ala}. It is the goal of this article to make a step into this direction.

\subsection*{The symmetry-restricted state sum}

In many calculations performed -- both analytically and numerically -- in lattice gauge theory, the art is to approximate the path integral in the right way. The goal is to make it simple enough to handle, while still keeping an expression from which physical information can be extracted. One way to perform this is to not sum over all histories of the fields in the path integral, but only over those which are relevant for the physical process one is interested in. That way, one can still hope to gain realistic approximations for the values of certain observables. Of course, the approximation in question therefore has to be tailored to the situation one wants to investigate.

We intend to copy this strategy for the spin foam approach in this article. We believe that this could, if done in the right way, provide very useful approximations for the spin foam state sum. Hence this might ultimately provide a pathway to statements not only about the continuum limit of spin foams, but about expectation values of physical observables, as well as the renormalization group flow of the model. In all of these, the spin foam state sum has to be performed, which in full generality can most likely not be achieved.

As a first step, in our article we will look at a drastic approximation of the whole path integral. These will render the physical predictability of the model questionable in certain regions. However, the model will still be complicated enough so that certain concepts can be tested, and specific open questions can be investigated.

In particular, we will work on a \emph{regular hypercubic lattice} in 4d. Furthermore, on this lattice, we will not consider all possible states, but only those which conform to the lattice symmetry. This is a condition on the intertwiners, which we pick to correspond to \emph{cuboids}. \footnote{This choice has also been considered in \cite{Alesci:2015nja} in the canonical context, and in \cite{Rennert:2013pfa, Rennert:2013qsa} in the cosmological setting. The hypercubic lattice has also been investigated in the loop quantum gravity path integral framework in \cite{Martins:2008zf, Baratin:2008du}.}

\begin{figure}[!h]
\includegraphics[width= 0.28\textwidth]{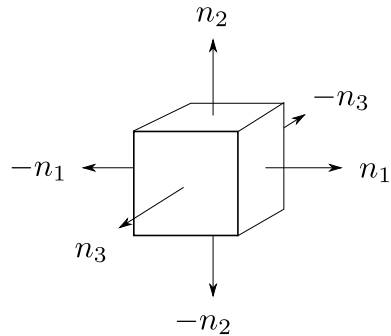}
\caption{A cuboid in $\mathbbm{R}^3$. The face normals $n_i$ are orthogonal, and normalized by the face areas.  \label{Fig:Cuboid01}}
\end{figure}

A cuboid is completely determined by its three edge lengths, or equivalently by its three areas (see figure \ref{Fig:Cuboid01}). All internal angles are $\frac{\pi}{2}$, and the condition of regular cuboids on all dual edges of the lattice will result in a high degree of symmetries on the labels: The area (and hence the spin) on each two parallel squares of the lattice which are translations perpendicular to the squares, have to be equal.

A few comments about this simplification are in order:
\begin{itemize}
\item The main reason for this symmetry is that it simplifies the problem dramatically, making the summation over states much more manageable. Also, the lattice is directly tailored for making contact with usual forms of lattice gauge theory, which we will explore by coupling matter degrees of freedom in a companion paper, in the spirit of \cite{Oriti:2002bn}.
\item As one can readily see, the high degree of symmetry will make all quantum geometries \emph{flat}. The analysis carried out in this article is therefore definitely not suited for describing local curvature. Still, there remain enough states to describe local degrees of freedom. These will be \emph{non-geometric}, as a result of the twisted geometries appearing in the spin foam formalism \cite{Freidel:2010aq, Freidel:2013bfa}. The role and behaviour of this non-geometricity will form a major part in this article.
\item The set of considered states is still large enough so that an infinite-dimensional (Abelian) subgroup of the diffeomorphisms acts on it. This makes this model an ideal test-bed for the question of how to treat these diffeomorphisms on the lattice, and how to treat them in the state-sum.
\item Restricting the state sum to only a subclass of states is not only a tool to simplify the state sum. It results in an approximation which can be valid for describing situations in which the removed states are not suspected to have a great influence on the physics. For instance, one could disregard states with high curvature if one is only interested in the low-curvature regime of the path integral. While considering only quantum cuboids is certainly a very severe restriction, one could try to access, say, the propagation of gravitational waves on a lattice, by adding very specific states describing low local curvature. A controlled enlargement of the states beyond the set of quantum cuboids could therefore provide an excellent tool for approximations of the spin foam path integral.
\end{itemize}

\noindent The plan of our article is as follows:

We will give a brief review of the EPRL-FK spin foam model in section \ref{Sec:Introduction}. Before performing computations in the quantum regime, we will look at the semiclassical regime of the path integral. The corresponding geometrical intuition, which we will develop in section \ref{Sec:Semiclassical}, will be very useful for the rest of the article.

The actual construction of the quantum cuboid intertwiner will be carried out in section \ref{Sec:QuantumCuboids}. Afterwards, the full vertex amplitude will be considered in section \ref{Sec:Amplitude}, where we in particular describe its asymptotic expression for large spins (i.e.~areas). Finally, the results of our numerical investigation of the quantum path integral for more than one building block, in particular in terms of the geometric interpretation developed earlier, will be given in section \ref{Sec:Results}.

We will see that, despite the severe simplification of the system by restricting to quantum cuboids, the analysis will reveal a surprising amount of insight into the path integral.

\section{Introduction}\label{Sec:Introduction}

The spin foam state sum we will employ in this article will be the Euclidean EPRL-FK model with Barbero-Immirzi parameter $\gamma<1$. The main reason for this is that this model is, at the current state, one of the best understood ones, with a lot of literature already existing on the subject.

The EPRL-FK model was originally defined on a 2-complex dual to a 4d triangulation, but there is a clear generalization to arbitrary 2-complexes \cite{Kaminski:2009fm}, which we will use in this article. A 2-complex $\Gamma$ is determined by its vertices $v$, its edges $e$ connecting two vertices, and faces $f$ which are bounded by the edges. While this data is purley combinatorial, we will think of $\Gamma$ as being embedded in a manifold. For practical purposes one needs to choose fiducial orientations of edges and faces, where however the result does not depend on that choice.

The spin foam model can either be written in an operator formulation \cite{Bahr:2010bs}, or in a dual holonomy formulation \cite{Bahr:2012qj}. In the former one, the path integral is formulated as a sum over \emph{states}. A state in this context is given by a collection of \emph{spins}, i.e.~irreducible representations $j_f\in\frac{1}{2}\mathbbm{N}$ of $SU(2)$ to the faces, as well as a collection of \emph{intertwiners} $\iota_e$ on edges. Here an intertwiner on the edge $e$ is an invariant tensor in the space $V_{j_1}\otimes V_{j_2}\otimes \ldots \otimes V_{j_n}$, where $j_1,\ldots j_n$ are the spin on the faces meeting at $e$, and the representation spaces $V_{j_i}$ are either the usual one or the dual representation space for that spin, depending on whether the relative orientations of $e$ and $f$ agree or disagree.

The actual sum is given by
\begin{eqnarray}\label{Eq:StateSum}
Z_\Gamma\;=\;\sum_{j_f,\iota_e}\prod_f\mathcal{A}_f\prod_e\mathcal{A}_e\prod_v\mathcal{A}_v,
\end{eqnarray}

\noindent where $\mathcal{A}_f$, $\mathcal{A}_e$ and $\mathcal{A}_v$ are the \emph{face}-, \emph{edge}- and \emph{vertex-amplitude} functions, depending on the state. The sum has to be carried out over all spins, and over an orthonormal basis in the intertwiner space at each edge. \footnote{There is a subtlety at this point: if one sees the state sum as coming from a restriction of the 4d BF amplitude on which the simplicity constraints have been imposed, one should actually not sum over an orthonormal basis of $SU(2)$-intertwiners, but rather over a set $\iota_e$ such that $\Phi(\iota_e)$ form an orthonormal subset of the $SU(2)\times SU(2)$-intertwiners, where $\Phi$ is the boosting map. These two choices are not the same, since it can be shown that $\Phi$ is not an isometry \cite{Kaminski:2009cc}. Since in this article, we will only have one specific intertwiner for each edge, the issue does in fact not arise. It will, however, emerge as soon as one allows for intertwiner fluctuating in the path integral.  }

The allowed spins $j_f$ in the EPRL-FK model are such that $j_f^{\pm}:=\frac{|1\pm\gamma|}{2}j_f$ both are also spins, i.e.~half-integers.\footnote{This means that, depending on $\gamma$, the set of allowed spins can actually be very small, or even just be the trivial representation, if $\gamma$ is irrational. This feature is very particular to the Euclidean version of the model, and disappears in the Lorentzian setting, where all real $\gamma$ are allowed.}

Popular choices\footnote{This choice has influence on the convergence of the state sum, see e.g.~\cite{Bonzom:2013ofa, Riello:2013bzw}} for the face amplitudes are either $\mathcal{A}_f=2j_f+1$ or $\mathcal{A}_f=(2j^+_f+1)(2j^-_f+1)$. In what follows, we keep an open mind about the precise form of the face amplitude, and allow for some freedom. We in particular introduce an additional parameter $\alpha$, and write
\begin{eqnarray}
\mathcal{A}_f=\Big((2j_f^++1)(2j_f^-+1)\Big)^\alpha,
\end{eqnarray}

\noindent and investigate the influence different choices for $\alpha$ have on the state sum.

The edge amplitudes $\mathcal{A}_e$ are usually taken to be equal to $1$ in our formulation, but there are certain ways of formulating the vertex amplitude, in which one puts the normalization of the intertwiners, or left over sign factors here.

The main ingredient to the state sum (\ref{Eq:StateSum}) is the vertex amplitude, the construction of which we briefly describe in what follows. For this we need to describe the boosting map $\Phi$, which maps the $SU(2)$-intertwiners $\iota_e$ to an $SU(2)\times SU(2)$-intertwiner by
\begin{eqnarray*}
\Phi\;:\;{\rm Inv}_{SU(2)}\bigotimes_iV_{j_i}\;\longrightarrow\;{\rm Inv}_{Spin(4)}\bigotimes_{i}V_{j_i^+}\otimes V_{j_i^-}.
\end{eqnarray*}

\noindent The map $\Phi$ is given by $\Phi=\mathcal{P}\circ\beta^{\otimes n}$, where $\mathcal{P}$ is the projector on $SU(2)\times SU(2)$-intertwiners, and the map $\beta: V_j\to V_{j^+}\otimes V_{j^-}$ is given, in the case that $\gamma<1$ which we use throughout this article, by the isometric embedding of $V_j$ into $V_{j^++j_-}$, the highest weight space in the Clebsh-Gordan decomposition of $V_{j^+}\otimes V_{j^-}$ in the $SU(2)$ representation category.

The vertex amplitude at the vertex $v$ is the contraction
\begin{eqnarray}\label{Eq:VertexDefinition}
\mathcal{A}_v\;=\;{\rm tr}\Big(\bigotimes_{e\supset v}\Phi(\iota_e)\Big)
\end{eqnarray}

\noindent where one takes either $\Phi(\iota_e)$ or the dual $(\Phi\iota_e)^\dag$, depending on whether the edge $e$ is incoming or outgoing of the vertex $v$. The ${\rm tr}$-contraction is to be understood in the sense that the tensor product in (\ref{Eq:VertexDefinition}), for each face $f$ touching $v$, there will be two $Spin(4)$ indices in the representation $(j_f^+,j_f^-)$ which can be contracted since they appear in opposite position.

\subsection{Coherent intertwiners}

A particular advancement in the spin foam model approach was the observation that the intertwiners $\iota_e$ can be given a geometric interpretation in terms of polyhedra in $\mathbbm{R}^3$ \cite{Livine:2007vk, Livine:2007ya, Bianchi:2010gc}. For a spin $j$, denote by $|j\,j\rangle$ the highest weight vector with respect to $\tau_3=\frac{i}{2}\sigma_3$, corresponding to the unit vector in the $3$- or $z$-direction. For any other unit vector $\vec n\in S^2$, choose $g\in SU(2)$ such that $g\triangleright e_3=\vec n$, then the coherent vector
\begin{eqnarray}
|j\,\vec n\rangle\;=\;g\triangleright|j\,j\rangle.
\end{eqnarray}

\noindent Given a collection of spins $j_1,\ldots j_n$ and vectors $\vec n_1,\ldots \vec n_n$ which close, i.e.~$\sum j_i\vec n_i=0$, one can define the \emph{coherent polyhedron}
\begin{eqnarray}\label{Eq:CoherentInterwiner}
|\iota\rangle\;=\;\int_{SU(2)}dg\;g\triangleright\bigotimes_i|j_i\,\vec n_i\rangle.
\end{eqnarray}

\noindent The geometric interpretation is that of a polyhedron, with face areas $j_f$ and face normals $\vec n_i$. The closure condition ensures that such a polyhedron exists, and is unique up to translation. The formula (\ref{Eq:CoherentInterwiner}) also exists for non-closing normals, which do not have a polyhedral interpretation. However, it can be shown that the state sum model (\ref{Eq:StateSum}) can be written solely as an integral over coherent polyhedra, i.e.~having closing normals \cite{Conrady:2009px}.\\[5pt]

This concludes the recap of the general setup. In the following we will describe the specific simplifications within this setup that we will use in order to obtain physical intuitions for the full path integral.

\section{Semiclassical considerations}\label{Sec:Semiclassical}

In this article, we will consider the case where the space-time manifold $\mathcal{M}\sim T^3\times[0,1]$ is the product of the 3-torus $T^3$ and a closed interval. Hence space is compactified toroidally. We cover $\mathcal{M}$ by $4d$ hypercubes, which form a regular hypercubic lattice $\mathbf{H}$. The 2-complex $\Gamma$ we use is the 2-skeleton of the dual complex. So there is a vertex for each hypercube, and two vertices are connected by an edge whenever two hypercubes intersect in a 3d cube. The faces of $\Gamma$ are dual to squares in $\mathbf{H}$, on which four hypercubes meet.

Note that this hypercubic lattice  does not carry any geometric information at this point. The geometry will later be encoded in the state, by specification of spins $j_f$ and intertwiners $\iota_e$.

In what follows, we consider the geometric interpretation of the semiclassical regime, not only of one intertwiner, but of the whole lattice. We are in particular interested in the large $j$-regime of the quantum cuboids. In this limit, these become classical cuboids as in figure \ref{Fig:Cuboid01}, which are completely specified by their three areas. The fact that opposite areas are identical lead to a translation symmetry in the areas of the lattice. Therefore, a  \emph{semiclassical configuration} is given by an assignment of areas $a = j\ell_P^2$ to the squares of the hypercubic lattice, where two areas agree whenever they are parallel and one is reached from the other by a translation perpendicular to them.

Denote the four directions in the lattice by $x, y, z, t$, and numerate the hypercubes by $\vec n\in\mathbbm{Z}^4$. The areas $a_{\mu\nu}^{\vec n}$ then satisfy
\begin{eqnarray}\label{Eq:LatticeSymmetry}
a_{\mu\nu}^{\vec n + p e_\rho+ q e_\sigma} = a_{\mu\nu}^{\vec n}
\end{eqnarray}

\noindent with $\mu,\nu,\rho,\sigma$ all different directions, $e_\rho,e_\sigma$ unit vectors in $\rho$, and $\sigma$ direction, and $p,q\in\mathbbm{Z}$.

Dual to one vertex $\vec n$ in the lattice is then a \emph{semiclassical quantum hypercuboid}, which is determined by the values of six of its areas $a_{xy}^{\vec n},a_{xz}^{\vec n},\ldots,a_{zt}^{\vec n}$ . Since a classical hypercuboid in $\mathbbm{R}^4$ is determined by its four edge lengths, this means there are two excess degrees of freedom. It is straightforward to see that these cannot correspond to geometric configurations: Assume that six areas are given, these determine uniquely the three areas for each individual 3d cuboid in the hypercuboid's boundary. In turn, for each such 3d cuboid, the three areas determine three edge lengths. In order for there to be a geometric hypercuboid, not only do the areas of two cuboids meeting in a rectangle have to match, but both individual edge lengths of that rectangle have to agree. It is not difficult to derive that the conditions the areas have to satisfy, so that both edge lengths of each rectangle, as seen from neighboring cuboids, agree. This condition can be written as

\begin{eqnarray}\label{Eq:GeometricityConstraints}
a_{xy}a_{zt}=a_{xz}a_{yt}=a_{xt}a_{yz}.
\end{eqnarray}

\noindent It seems worthwhile noting that this problem -- of matching areas but non-matching shape -- is well-known, and usually referred to as twisted geometries in the literature \cite{Freidel:2010aq, Freidel:2013bfa}.

The $4$-volume of a geometric hypercube is given by the product of areas of two perpendicular faces, so the geometricity constraints (\ref{Eq:GeometricityConstraints}) are the condition that the three choices of opposite pairs among the six faces deliver the same result. For a non-geometric configuration, we define the 4-volume of a hypercube as
\begin{eqnarray}\label{Eq:4Volume}
V_4\;:=\;\big(a_{xy}a_{xz}\cdots a_{zt}\big)^{\frac{1}{3}},
\end{eqnarray}

\noindent which reduces to the original definition at geometricity. If one were to define the four diameters to be
\begin{eqnarray}
d_x\;:=\;\frac{V_4}{(a_{yz}a_{yt}a_{tz})^{\frac{1}{2}}}\;=\;\frac{(a_{xy}a_{xz}a_{xt})^{\frac{1}{3}}}
{(a_{yz}a_{yt}a_{tz})^{\frac{1}{6}}}
\end{eqnarray}

\noindent with similar definitions for the other three directions in the lattice, one would have that $V_4=d_xd_yd_zd_t$, even for non-geometric configurations. We also define the \emph{non-geometricity}
\begin{eqnarray}\label{Eq:NonGeometricity}
\xi^{\vec n}\;:=\;\left(\begin{array}{c}
a_{xy}a_{zt}-a_{xz}a_{yt}\\
a_{xz}a_{yt}-a_{xt}a_{yz}\\
a_{xy}a_{zt}-a_{xt}a_{yz}
\end{array}\right)
\end{eqnarray}

\noindent as a measure of the deviation from the constraints (\ref{Eq:GeometricityConstraints}).

\subsection{Lattice deformations}

The fact that the hypercubic lattice in the manifold is not prescribed with an ab initio geometry, but obtains geometric interpretation via the state, has important consequences. In particular, there are some configurations which arise as lattice deformations of one another. These can be seen as the result of either an active, or a passive diffeomorphism.

\begin{figure}[!h]
\includegraphics[width= 0.48\textwidth]{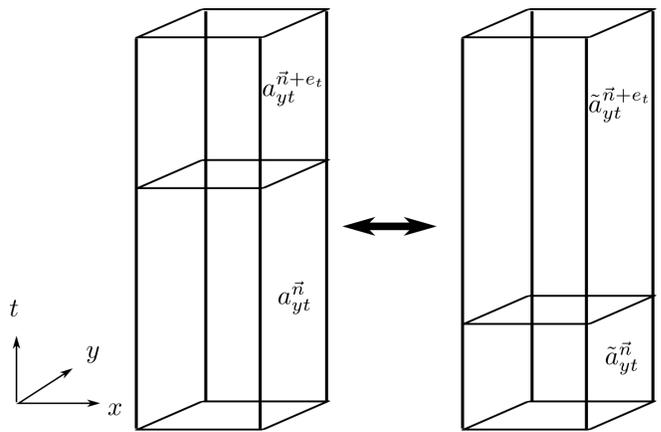}
\caption{Action of a diffeomorphism in $t$-direction on the areas $a_{it}^{\vec n}$ \label{Fig:QuasiDiffeomorphism01}}
\end{figure}

Imagine, for definiteness, the manifold to be equipped with a standard, flat Euclidean metric $\eta$, and the lattice $\Gamma$ inserted being completely regular, i.e.~all lengths (and hence all areas) are equal. Now consider a diffeomorphism $\phi$ which deforms the metric in a thin 3d strip, say, the slightly thickened hyperplane around $t=0$. Assume that $\phi$ is such that it stretches space below, and contracts the space above the hyperplane. As a result, with respect to the new metric $g=(\phi^{-1})^*\eta$, the areas of the ``time-like'' squares above the hyperplane will be smaller, and the ones below it will be larger (see figure \ref{Fig:QuasiDiffeomorphism01}).

So this diffeomorphism of the manifold $\mathcal{M}$ induces a change in the areas of the lattice, since it changes the metric, but leaves the lattice where it is (if it were to transform the lattice in a similar way, the areas would remain unchanged trivially). This can be regarded as an ``active'' diffeomorphism -- the "passive" version can be seen as keeping the metric fixed, but changing the embedding of the lattice $\Gamma\to\phi(\Gamma)$, by moving the $t=0$-hyperplane into positive $t$-direction. The result on the values of the areas in the lattice will be the same, so by just looking at the lattice and not at the manifold, the difference between ``active'' and ``passive'' diffeomorphism vanishes.

The continuum diffeomorphisms prescribed in this way are generated by vectorfields which only act in $t$ direction on the manifold, i.e.~$X(x, y, z, t)=T(t)\partial_t$, with $T$  a smooth function. Together with the deformations in the other three major directions, the resulting collection of diffeomorphisms forms an Abelian subgroup of the full group of ${\rm Diff}(\mathcal{M})$.

The variable transformation we have just described can be extended to the possibly non-geometric configurations that arise in the large-$j$-regime of the EPRL-FK model. Just as in the example, there is one generator of the group for each 3d hypersurface in the lattice, ``moving'' that hypersurface in a direction orthogonal to it, while keeping the rest of the lattice fixed. As one can see, this group is finite-dimensional for compact manifolds, but the dimension grows as the number of hypercubes increases, becoming infinite in the limit of infinitely large lattices.

To describe the action of these ``lattice deformations'' on the areas $a^{\vec n}_{\mu\nu}$, consider, for definiteness, a 3d hypersurface in the $xyz$-directions. Let $\vec n$ be a hypercube whose future $xyz$ cuboid is part of the chosen hypersurface, then the transformation is defined to be

\begin{eqnarray}
a_{it}^{\vec n}\to a_{it}^{\vec n}(1+\Delta^{\vec n})\qquad i=x, y, z
\end{eqnarray}

\noindent for all such $\vec n$, while $a_{ij}^{\vec n}$ remains unchanged. The time-like areas of the following hypersurface gets changed as well, by
\begin{eqnarray}
a_{it}^{\vec n+e_t}\to a_{it}^{\vec n+e_t}(1-\Delta^{\vec n+e_t})\qquad i=x, y, z.
\end{eqnarray}

\noindent The parameters $\Delta^{\vec n}$, $\Delta^{\vec n+e_t}$ quantify the stretching and compressing of the respective hypersurfaces. It is easy to see that, in order not to violate (\ref{Eq:LatticeSymmetry}), $\Delta^{\vec n}$ has to be the same among the whole hypersurface, i.e.~$\Delta^{\vec n+m_xe_x+m_ye_y+m_ze_z}=\Delta   ^{\vec n}=:\Delta_1$ for all $m_i$. A similar statement holds for $\Delta_2:=\Delta^{\vec n+e_t}$.

\begin{figure}[!h]
\includegraphics[width= 0.2\textwidth]{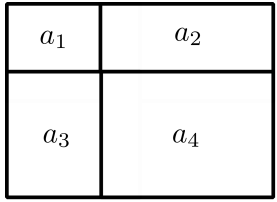}
\caption{The areas of flat squares in Euclidean space arranged in a tile satisfy $a_1a_4=a_2a_3$ \label{Fig:QuasiDiffeomorphism03}}
\end{figure}

There is an interesting observation to be made at this point: In general a diffeomorphism should not change the total 4-volume of space-time. Depending on the areas $a_{\mu\nu}^{\vec n}$ there might be no choice for $\Delta_{1,2}$ so that the 4-volume is preserved, however. So, while the action of the diffeo group can be defined on all configurations, there are only some on which it can be defined in a 4-volume-preserving way. We call these configurations \emph{semi-geometric}. A sufficient condition for semi-geometricity is that the sum of the 4-volumes of a pair of hypercubes $\vec n$, $\vec n+e_t$ stays constant under a diffeomorphism, so 4-volume is not only preserved in general, but also locally. It is not difficult to show that this is equivalent to the following condition: For every $2\times 2$-arrangement of squares in the lattice, the two respective products of diagonally opposite areas in that arrangement are equal (see figure \ref{Fig:QuasiDiffeomorphism03}). It should be clear that the diffeomorphisms, as we have defined them here, preserve the notion of semi-geometricity.

It should also be noted that semi-geometricity is weaker than geometricity (\ref{Eq:GeometricityConstraints}): a totally translation-symmetric configuration where all hypercubes are congruent, is certainly semi-geometric. However, the six areas of the hypercube can be chosen in a non-geometric way. Note that geometricity is purely local, in terms of one hypercube, while semi-geometricity is a condition on relations of areas between neighboring hypercubes.

\section{Quantum cuboids}\label{Sec:QuantumCuboids}

We now turn to the quantum theory.

In the 2-complex we consider, every edge has six faces attached to it, corresponding to the six faces of the cubes. So any intertwiner in the state-sum will be six-valent, and therefore can be described by a coherent polyhedron with six faces. In our setup, we restrict the state-sum to coherent cuboids, or \emph{quantum cuboids}. A cuboid is characterized by areas on opposite sides of the cuboid being equal, and the respective normals being negatives of one another (see figure \ref{Fig:Cuboid01}).

The state $\iota_{j_1, j_2, j_3}$ is given by
\begin{eqnarray}\label{Eq:QuantumCuboid}
|\iota_{j_1,j_2,j_3}\rangle\;=\;\int_{SU(2)}dg\;g\triangleright\bigotimes_{i=1}^3|j_i,e_i\rangle|j_i,-e_i\rangle.
\end{eqnarray}

\noindent Here $e_1=\exp(-i\pi\sigma_2/4)\triangleright e_3$, $e_2=\exp(i\pi\sigma_1/4)\triangleright e_3$, and $e_3$ are taken to be unit vectors in $\mathbbm{R}^3$. It is worth noting that the intertwiner space is nonempty for every choice of spins $j_1, j_2, j_3$, and the state (\ref{Eq:QuantumCuboid}) always exists and is non-vanishing.

\section{The amplitude}\label{Sec:Amplitude}
\begin{figure}[h]
\includegraphics[width= 0.48\textwidth]{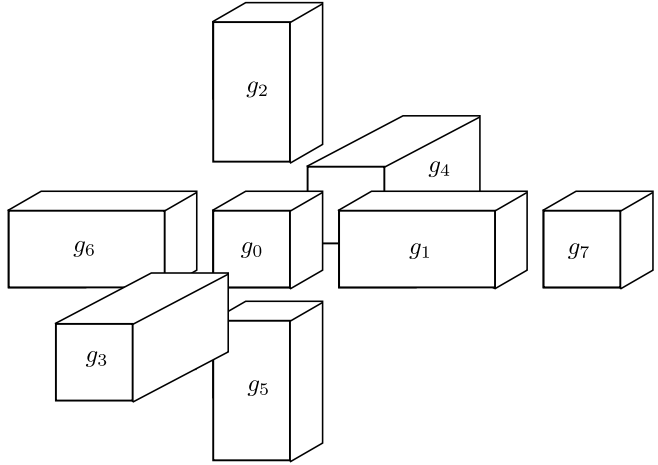}
\caption{A four-dimensional hypercuboid represented by its boundary, consisting of eight cuboids. They are labelled by $a=0\ldots 7$, as depicted.  \label{Fig:Hypercube01}}
\end{figure}

\begin{figure}[hbt!]
\begin{center}
\includegraphics[width= 0.48\textwidth]{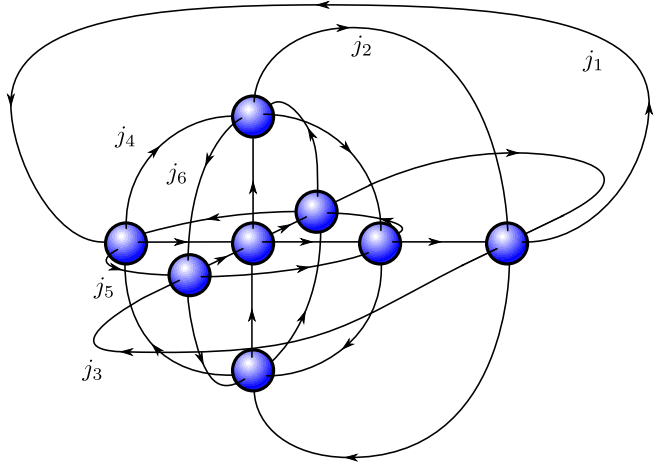}
\end{center}
\caption{The spin network for a hypercuboid. Each node is six valent, and spins of edges emanating opposite of each other from a node coincide. Therefore, only six different spins exist. In the semiclassical limit these become the six areas $a_{\mu\nu}$ of the hypercuboid. \label{Fig:Hypercube02}}
\end{figure}

 The vertices in the four-dimensional hypercubic lattice are all eight-valent, since a 4d hypercube is bounded by eight cuboids (see figure \ref{Fig:Hypercube01}). The vertex amplitude is defined with the help of a boundary spin network $\Gamma$, which in our case is of the special form depicted in figure \ref{Fig:Hypercube02} . The amplitude is given in terms of spins $j_l$ associated to links $l$ of $\Gamma$, as well as coherent intertwiners $\iota_a$ (\ref{Eq:CoherentInterwiner}) associated to the nodes $a$ of $\Gamma$. The intertwiner can be given in terms of normal vectors $\vec n_{ab}$, which is thought of being the normal vector of the link $l=(ab)$, sitting at the polyhedron at $a$.

The vertex amplitude for a Barbero-Immirzi parameter $\gamma<1$ factorizes as $\mathcal{A}_v=\mathcal{A}_v^+\mathcal{A}_v^-$, with
\begin{eqnarray}\label{Eq:AmplitudeIntegral}
\mathcal{A}_v^\pm = \int_{SU(2)^N}dg_a\;e^{S^\pm[g_a]}
\end{eqnarray}

\noindent with the complex action
\begin{eqnarray}\label{Eq:ComplexAction}
S^\pm[g_a]\;&=&\;\frac{1\pm\gamma}{2}\sum_l2j_l\ln\langle-\vec n_{ab}|g_{a}^{-1}g_{b}|\vec n_{ba}\rangle\\[5pt]\nonumber%
&=:&\;\frac{1\pm\gamma}{2}S[g_c]
\end{eqnarray}

\noindent where, in formula (\ref{Eq:ComplexAction}), $a$ is the source node of the link $l$, while $b$ is its target node, and all states are in the fundamental representation.

\subsection{Large $j$ asymptotics}

The amplitudes $\mathcal{A}_v^\pm$ possess an asymptotic expression for large $j_l$, which has been investigated in \cite{Barrett:2009gg, Barrett:2009mw}. To compute this expression, we follow the analysis in \cite{Barrett:2009gg}, and refer to that article for calculational details.

Firstly, we note that out of the $N=8$ group integrations in (\ref{Eq:AmplitudeIntegral}), one is obsolete due to the invariances of the Haar measure $dg$. The remaining integral possesses isolated critical points, and can therefore be evaluated by an extended stationary phase approximation. The action (\ref{Eq:ComplexAction}) is readily seen to be invariant under any $g_a\to -g_a$, so there is a $2^7$-fold symmetry.

Modulo this symmetry, there are two distinct stationary and critical points, satisfying the equations
\begin{eqnarray}\label{Eq:SolutionVector}
\tilde n_{ab}\;:=\;g_a\triangleright \vec n_{ab}= - g_b\triangleright \vec n_{ba}
\end{eqnarray}

\noindent for all links $(ab)$ (the closure condition is automatically satisfied by the choice of quantum cuboids, for any selection of spin $j_l$). Using the convention in figure \ref{Fig:Hypercube01}, having fixed $g_0=\mathbbm{1}$, the two solutions $\Sigma_1$ and $\Sigma_2$ are shown in table \ref{Tab:TwoSolutions}.\\

\begin{table}[hbt!]
  \centering
  \begin{tabular}{c|c|c}
& $\;\;\Sigma_1\;\;$ & $\;\;\Sigma_2\;\;$\\[5pt]\hline
$\;g_1\;$ &  $\;\exp\left(i\frac{\pi}{4}\sigma_1\right)\;$ & $\;\exp\left(-i\frac{\pi}{4}\sigma_1\right)\,$\\[5pt]
$g_2$ & $\exp\left(i\frac{\pi}{4}\sigma_2\right)$ & $\exp\left(-i\frac{\pi}{4}\sigma_2\right)$\\[5pt]
$g_3$ & $\exp\left(i\frac{\pi}{4}\sigma_3\right)$ & $\exp\left(-i\frac{\pi}{4}\sigma_3\right)$\\[5pt]
$g_4$ & $\;\exp\left(-i\frac{\pi}{4}\sigma_3\right)\;$ & $\exp\left(i\frac{\pi}{4}\sigma_3\right)$\\[5pt]
$g_5$ & $\exp\left(-i\frac{\pi}{4}\sigma_2\right)$ & $\exp\left(i\frac{\pi}{4}\sigma_2\right)$\\[5pt]
$g_6$ & $\exp\left(-i\frac{\pi}{4}\sigma_1\right)$ & $\exp\left(i\frac{\pi}{4}\sigma_1\right)$\\[5pt]
$g_7$ & $\exp\left(i\frac{\pi}{2}\sigma_1\right)$ & $\exp\left(i\frac{\pi}{2}\sigma_1\right)$

\end{tabular}
\caption{The two solutions for $g_1,\ldots g_7$ of (\ref{Eq:SolutionVector}) not connected by symmetry, corresponding to the stationary and critical points of the action (\ref{Eq:ComplexAction}). Note that $g_1,\ldots, g_6$ all correspond to rotations by $\pi/2$, while $g_7$ corresponds to rotations by $\pi$.}\label{Tab:TwoSolutions}

\end{table}

\noindent It can be easily checked that these are, in fact solutions to the extended stationary phase approximation equations (\ref{Eq:SolutionVector}). Also, one can readily see that these are the only solutions: the six equations for the edges $(01)$ up to $(06)$ force each of the respective six group elements $g_i$ $(i=1,\ldots 6)$ to lie in some $1$-parameter subgroup, corresponding to a rotation around the respective normal $\vec n_{0i}=\tilde{n}_{0i}$ of cuboid $0$. These angles are fixed by the equations (\ref{Eq:SolutionVector}) for $a,b\in{1,\ldots 6}$, and there are two overall solutions (modulo $2^6$ signs). Remaining are six vector equations for one $SU(2)$ element $g_7$, so the system appears drastically overdetermined. However, one finds that the solution $g_7=\pm\exp\left(i\frac{\pi}{2}\sigma_1\right)$ solves all of these simultaneously, and these are the only ones which do so. This leaves us with two solutions (up to $2^7$ signs), depicted in table \ref{Tab:TwoSolutions}.

Another comment about these solutions are in order: As in the case for the 4-simplex, there are ``non-geometric'' boundary data, which cannot be used to define a continuous $3$-metric on all of the boundary of the hypercuboid (in the 4-simplex case, these are called ``non-Regge-like''). \emph{Unlike} in the 4-simplex case, however, not all of these non-geometric boundary data are suppressed by the stationary phase approximation method. Rather, there are non-degenerate, but non-Regge-like states which appear in the large $j$-limit. These are described in section \ref{Sec:Semiclassical}, as states violating the geometricity condition (\ref{Eq:GeometricityConstraints}). This issue can be traced back to the differences in geometry of a 4-simplex and a 4-hypercuboid, in particular the fact that specifying all areas (satisfying certain inequialities) in the former determines a geometric state, while it does not in the latter. This point certainly deserved more attention, and we will come back to it at another point\footnote{We thank the anonymous referee who pointed out the importance of this issue.}.

Each critical stationary point $\vec g_{c}$ contributes one term of the form
\begin{eqnarray}
I\;=\;\sqrt{\frac{(2\pi)^{21}}{\det (-H(\vec g_c))}}\;e^{S(\vec g_c)}
\end{eqnarray}
\noindent to the approximation formula, where $H(\vec g_c)$ is the Hessian matrix of $S$, evaluated on the points $\vec g=\vec g_c$.

One can readily see that both critical stationary points listed in table \ref{Tab:TwoSolutions} lead to $S[\vec g_c]=0$. This is not surprising, since all dihedral angles equal $\frac{\pi}{2}$, and at each square in the lattice precisely four hypercuboids meet, leading to vanishing deficit angle. In other words, the choice of quantum cuboids as the only allowed intertwiners prevents local curvature (intrinsic and extrinsic) to be excited. Hence the Einstein-Hilbert action part vanishes, as well as the boundary terms.

The stationary phase approximation for the amplitude is therefore completely determined by the determinant of the Hessian matrix $H$ of second derivatives of $S$. This determines the (first order) quantum corrections to the classical solution in the path integral. It therefore contains nontrivial quantum information about the path integral measure, beyond the Einstein-Hilbert action.

Introducing, around each critical point $\vec g_c$, coordinates $\{X_a^I\}$ via $g_a=g_{a,c}\exp(i\sigma _I X_a^I/2)$, one obtains for the entries of the Hessian matrix
\begin{eqnarray}
\frac{\partial^2S}{\partial X_a^I\partial X_a ^J}\;&=&\;-\sum_{(ab)\supset a}\frac{j_{ab}}{2}\Big(\delta_{IJ}-\tilde n_{ab}^I\tilde n_{ab}^J\Big)\\[5pt]\label{Eq:OffDiagonalHessian}
\frac{\partial^2S}{\partial X_a^I\partial X_b ^J}\;&=&\;\frac{j_{ab}}{2}\big(\delta_{IJ}-i\epsilon_{IJK}\tilde n_{ab}^K-\tilde n_{ab}^I\tilde n_{ab}^J\big)
\end{eqnarray}

\noindent where the vectors $\tilde n$ are given by (\ref{Eq:SolutionVector}). Note that the second expression (\ref{Eq:OffDiagonalHessian}) is symmetric, because $\tilde n_{ab}=-\tilde n_{ba}$. The Hessian matrix is $21\times 21$, and its determinant can be computed by a computational algebra program. The result is a quite long expression, which can nevertheless be handled analytically and numerically. As it turns out, the two determinants for $\Sigma_1$ and $\Sigma_2$ are complex conjugate of each other, so that the whole amplitude is real.

It is worth noting at this point that, because $S=0$ on critical stationary points for the quantum hypercuboid, the only dependence on the Barbero-Immirzi parameter $\gamma$ is via the Hessian matrix determinant $\det H$. Since $\det H(\lambda j)=\lambda^{21}\det H(j)$, i.e.~$H$ is homogenous in the spins, one can readily see that the amplitudes $\mathcal{A}_v^\pm$ satisfy, in the large $j$ limit,
\begin{eqnarray}
\tilde{\mathcal{A}}_v\;:=\;\left(\frac{1+\gamma}{2}\right)^\frac{21}{2}\mathcal{A}_v^+\;=\;\left(\frac{1-\gamma}{2}\right)^\frac{21}{2}\mathcal{A}_v^-.
\end{eqnarray}

\noindent Similarly, in the large $j$-limit, the norm squared of the quantum cuboid states (\ref{Eq:QuantumCuboid}) is given by
\begin{eqnarray}
\|\Phi\iota_{j_1j_2j_3}\|^2\;\sim\;\frac{8(1-\gamma^2)^{-\frac{3}{2}}}{(j_1+j_2)(j_2+j_3)(j_1+j_3)}
\end{eqnarray}

\noindent With this, we get for the state sum, in the large-$j$ limit on a regular hypercubic lattice, that
\begin{widetext}
\begin{eqnarray}\nonumber
Z\;&\sim&\;\left(\frac{1-\gamma^2}{4}\right)^{\alpha F-\frac{3}{2}E+\frac{21}{2}V}\sum_{j_f}\prod_fj_f^{2\alpha}\prod_e(j_1+j_2)(j_2+j_3)(j_1+j_3)\prod_v\tilde{\mathcal{A}}_v^2\\[5pt]\label{Eq:AsymptoticStateSum}
&=:&\;\left(\frac{1-\gamma^2}{4}\right)^{(6\alpha-9/2)V}\sum_{j_f}\prod_v\widehat{\mathcal{A}}_v.
\end{eqnarray}
\end{widetext}

\noindent In the last line, we have defined the dressed vertex amplitude $\widehat{\mathcal{A}}_v$, in which the face- and edge amplitudes have been absorbed in such a way that boundary amplitudes are taken care of correctly. Note that the only way the asymptotic state sum (\ref{Eq:AsymptoticStateSum}) depends on the Barbero-Immirzi paramter $\gamma$ is via the prefactor, which does not influence the physics. This is a direct consequence of the fact that no curvature degrees of freedom are excited in our simplified state sum.

\subsection{Properties of the amplitude}

For the following article, we will use the following notation: the (dressed) vertex amplitude $\widehat{\mathcal{A}}_v$ depends on six spins $j_1,\ldots, j_6$, so we write
\begin{eqnarray}
\widehat{\mathcal{A}}_v(\vec j)\;=\;\widehat{\mathcal{A}}_v(j_1,j_2,j_3, j_4, j_5, j_6),
\end{eqnarray}

\noindent using the notation in figure \ref{Fig:Hypercube02} . In particular, $j_{1,2,3}$ denote the spins on space-like, $j_{4,5,6}$ the spins on time-like faces. On some occasions in what follows, we will treat space- and time-like faces separately, so we also introduce the short-hand
\begin{eqnarray}\label{Eq:ShorthandNotation01}
\widehat{\mathcal{A}}_v(\vec j, \vec k)\;:=\;\widehat{\mathcal{A}}_v(j_1, j_2, j_3, k_1, k_2, k_2),
\end{eqnarray}

\noindent for the amplitude of the equilateral hypercuboid. So whenever the amplitude has two arguments, they refer to space- and time-like spins, respectively.

In the hypercubic lattice the relation $V=\frac{F}{6}=\frac{E}{4}$ between the numbers of vertices, faces and edges, holds. A term $\widehat{\mathcal{A}}_v$ in the state sum therefore scales with the spins as
\begin{eqnarray}
\widehat{\mathcal{A}}_v(\lambda \vec j)\;\sim\;\lambda^{12\alpha-9}.
\end{eqnarray}
\noindent This means that for a critical value of $\alpha_{\infty}=\frac{3}{4}$, the amplitude becomes scale-invariant. For larger $\alpha$, the state sum is most definitely divergent, while there can be found a value of $\alpha$ small enough so that the state sum converges.

The situation is different if we do not consider the state sum for a closed manifold, but one with boundary. Consider a finite regular hypercubic lattice decomposition of $\mathcal{M}=T^3\times [0,1]$. The spatial boundary manifolds are three-tori, and fixing boundary spin networks is equivalent to fixing all spatial spins throughout the lattice, due to the symmetry (\ref{Eq:LatticeSymmetry}). The scaling of an amplitude with respect to the time-like spins goes like
\begin{eqnarray}\label{Eq:ScalingInTimeDirection}
\widehat{\mathcal{A}}_v(\vec j,\lambda \vec k)\;\sim\;\lambda^{6\alpha-9}.
\end{eqnarray}

\noindent This means that, when computing the physical inner product with the help of the state sum (\ref{Eq:AsymptoticStateSum}) on a hypercubic lattice, one can expect the sum to converge absolutely for $\alpha<\frac{3}{2}$, when the lattice is large enough.

\section{Results}\label{Sec:Results}

In the following, we will investigate the asymptotic amplitude for the hypercuboid. This will give an insight into the quantitative and qualitative behaviour of the spin foam state sum.

\subsection{Non-geometricity in the path integral}

One interesting question is how much configurations with high non-geometricity $\xi$ (\ref{Eq:NonGeometricity}) contribute to the path integral (\ref{Eq:AsymptoticStateSum}). To investigate this, we consider a regular geometric hypercube, i.e.~with all spins $a_{\mu\nu}=j\ell_P^2$ equal. Then we consider the value of the amplitude $\widehat{\mathcal{A}}_v$, depending on the two independent non-geometric directions of configurations $\xi$ around the regular configuration. We vary the three time-like areas $a_{it}$, $i=x, y, z$, by
\begin{eqnarray}
\delta\xi_1\;&=&\;\frac{1}{\sqrt{2}}\big(\delta a_{xt}-\delta a_{yt}\big)\\[5pt]\nonumber
\delta \xi_2\;&=&\;\frac{1}{\sqrt{6}}\big(\delta a_{xt}+\delta a_{yt}-2\delta a_{zt}\big).
\end{eqnarray}

\begin{figure}[hbt!]
\begin{center}
\includegraphics[width=0.48\textwidth]{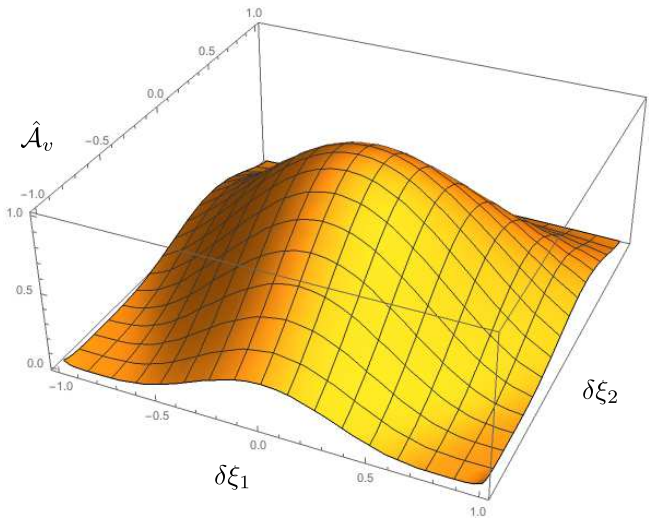}
\end{center}
\caption{Value of the dressed vertex amplitude depending on the non-geometricity of the labels.\label{Fig:NonGeometricity01}}
\end{figure}

\noindent Note that this variation is isochoric, i.e.~satisfies $\delta V=0$. The value of the amplitude is shown in figure \ref{Fig:NonGeometricity01}.

The amplitude shows a clear maximum at the geometric configuration. It is noteworthy that this is not a local maximum in all of state space, though. The reason is that there are four more directions in which the amplitude can be varied, which can change the shape of the hypercube. These directions preserve geometricity. Of those will be one direction corresponding to simply scaling the hypercube. Since the amplitude scales as (\ref{Eq:ScalingInTimeDirection}) with respect to this direction, the statement can therefore be phrased as follows: Among all configurations with a fixed four-volume, the ones contributing most to the path integral will be the ones with vanishing non-geometricity.

The analysis suggests that fluctuations in the non-geometricity $\xi$ around $\xi=0$ do contribute to the path integral, proportional to the width of the Gaussian in figure \ref{Fig:NonGeometricity01}. This width $m_\xi^2$ can be interpreted as a mass term for $\xi$, since it measures the ease with which $\xi$ can be excited locally. The precise value of this mass depends on the parameter $\alpha$ in the path integral. The connection between the two can readily be computed from the exact form of the large-$j$ asymptotic formula, to be

\begin{eqnarray}
m_\xi^2(\alpha)\;=2\alpha\,-\,\frac{233}{240}.
\end{eqnarray}

\noindent so for $\alpha<\frac{233}{480}\approx 0.4854$, the mass $m_\xi^2<0$ will be negative. This will result in the main contribution to the path integral coming from highly non-geometric configurations. One could argue that this restricts the desired value of $\alpha$ on physical grounds, since the path integral should deliver geometric states in the classical limit.\\[5pt]

The presence of a mass term suggests that there could also be a kinematical term for $\xi$, turning the non-geometricity into a propagating degree of freedom. To investigate this further, one would not only use more than one building block, one would in particular have to consider more intertwiners than just the quantum cuboid, to investigate realistic propagation of local excitations of $\xi$.

In particular, there is the possibility that there exists a regime in the full path integral in which the non-geometricity $\xi$ becomes effectively described by a fluctuating quantum field on some background geometry. This point warrants further investigation.

\subsection{Vertex-displacement symmetry and the amplitude}

We now come to an investigation of the amplitude for more than one building block. To this end, we first consider two hypercubes, which are glued together along a common (space-like) cuboid. One hypercuboid can therefore be regarded as being ``in the future'' of the other. We assume that the main contribution comes from the geometric configurations, i.e.~$\alpha>0.4854$.

We consider the transition between two spin network functions. Because of the chosen symmetry (\ref{Eq:LatticeSymmetry}) within our lattice, this fixes all the spatial spins (and hence the initial and final spin networks to be equal). The only fluctuating spins are, therefore, the timelike ones, i.e. three per hypercuboid, six in total. Out of these three degrees of freedom per hypercube, there are two which are non-geometric. We disregard these in the following and observe that only two degrees of freedom remain in total for the two hypercuboids. While one is obviously related to the total $4$-volume, the other corresponds to a (generalized) diffeomorphism direction connected to a time-like translation of the bulk Cauchy hypersurface (as in  figure \ref{Fig:QuasiDiffeomorphism01}). In other words, when computing the physical inner product between two spin networks, using the 2-complex described above, the sum over spins will, essentially, be the sum over two parameters, corresponding to the total 4-volume of the universe, and a (quasi-)diffeomorphism degree of freedom.

\begin{figure}[hbt!]
\begin{center}
\includegraphics[width=0.48\textwidth]{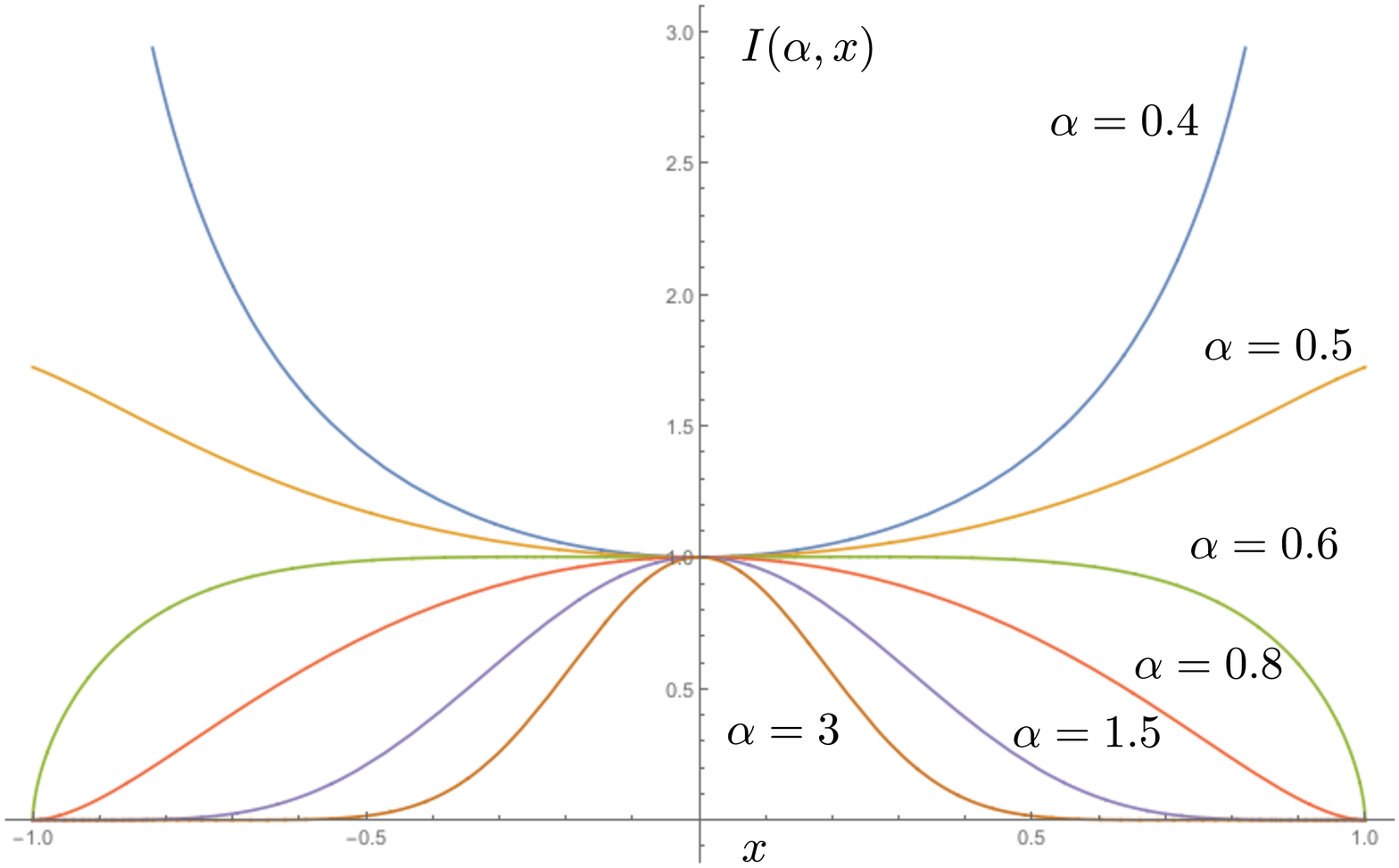}
\end{center}
\caption{Value of the product of two amplitudes $I(\alpha, x)$ depending on vertex translation $x$, for different values of the coupling constant $\alpha$. \label{Fig:Diffeomorphism02}}
\end{figure}

To shed light on the behavior of the amplitude with respect to this diffeomorphism, we consider the total amplitude for the whole lattice, which is the product of two vertex amplitudes. The spatial spins are all set to some large spin $j$, while the time-like spin of the two hypercuboids are set to $j(1+x)$ and $j(1-x)$, respectively. The amplitudes are normalized for $x=0$. The findings are shown in figure \ref{Fig:Diffeomorphism02}, for different values of $\alpha$.

\noindent Due to symmetry, it is clear that the derivative of the total amplitude $I(\alpha, x)=\widehat{\mathcal{A}}_{v}(j, j(1+x))\widehat{\mathcal{A}}_{v}(j, j(1-x))$ is zero at $x=0$. The second derivative depends linearly on $\alpha$, and obeys
\begin{eqnarray}
I''(\alpha, 0)\;=\,\frac{182}{25}\,-\,12\alpha.
\end{eqnarray}

\noindent This can be rephrased the following way: The contributions from different points along the diffeomorphism orbit parametrized by $x$ are different, and depend on $\alpha$. There is a critical value $\alpha_c=\frac{182}{300}\approx 0.6067$, and for $\alpha>\alpha_c$, the main contribution to the path integral comes from the point $x=0$, i.e.~the point where both hypercuboids are regular and of equal size. Conversely, for $\alpha<\alpha_c$, the main contribution to the path integral comes from very irregular configurations, i.e.~the points $x=\pm1$. Note that in the case of the path integral for more than two hypercuboids, the situation will be even more pronounced, because the total amplitude factorizes over the vertices.

One can therefore assert the following: the parameter $\alpha$ distinguishes between two different regions, in which the path integral behaves quantitatively different. For $\alpha>\alpha_c$, the main contribution to the path integral, among all geometric configurations of equal volume, comes from the \emph{completely regular} configuration, i.e.~all spins, and therefore all hypercuboids, being equal. This can be interpreted as very regular geometry, approximating $4$-dimensional Euclidean space.

On the other hand, for $\alpha<\alpha_c$, the main contribution to the path integral (again, among all geometric configurations of equal $4$-volume) come from configurations in which almost all $4$-volume is concentrated in one time-slice, while all other time-slices contain nearly none of it. These configurations can be interpreted as very irregular.

A comment is in order here: While this configuration can be regarded as higly irregular, it is nevertheless flat, and describes \emph{the same} classical 4-d metric as the one in which all $j$'s are equal, i.e. the completely regular case. Although the $j$-values are different, these two states can be regarded as diffeomorphically equivalent, since they are results of a vertex translation symmetry as described in section \ref{Sec:Semiclassical}. Geometrically, they correspond to different ways in which a 4-torus can be cut up into two hypercuboids -- one in which the two hypercuboids have the same size, and one in which one contains much more volume than the other.

In other words, we have a clear example of a coupling constant $\alpha$, which distinguishes between two regions in the phase diagram in which the path integral behaves qualitatively very different. Furthermore, on the critical value $\alpha=\alpha_c$, the amplitude function is nearly invariant under change of $x$, which one can readily interpret as invariance under vertex translation symmetry, or diffeomorphism-invariance. This fosters the scenario for this point as second order phase transition: one can assume that correlation lengths become infinite (in terms of lattice distance) because of diffeomorphism invariance. The reason for this is that close and far separation on the lattice become diffeomorphically equivalent (see also discussion in \cite{Dittrich:2012qb, Bahr:2014qza}). Of course, this point deserves much more investigation, in order to make this correspondence more precise.

One should note that the precise numerical value of $\alpha_c$ should not be taken too seriously at this point: it rests on the simplifications we have employed, as well as taking all spatial spins to be equal. For the spatial spins different, one finds qualitatively similar results, however, with slightly changed values of $\alpha_c$.

\section{Physical inner product and torsion}
\subsection{The role of the physical norm}
The path integral can be used to compute the physical inner product between states, and in particular to compute the physical norm of kinematical states. More precisely, the path integral functions as a rigging map
\begin{eqnarray}
\eta\;:\;\mathcal{H}_{\rm kin}\supset\mathcal{D}_{\rm kin}\;\longrightarrow\;\mathcal{D}_{\rm phys}\subset\mathcal{H}_{\rm phys}
\end{eqnarray}
\noindent from (a dense subset of) the kinematical to (a dense subset of) the physical Hilbert space. Usually, $\mathcal{D}_{\rm kin}$ is taken to be the linear span of the spin network functions. Different spin network functions (which are all normalized in the kinematical inner product) will have images under the rigging map $\eta$ with different norm in $\mathcal{H}_{\rm phys}$.

If a spin network state $\phi$ on the (say, initial) boundary is given by spins $j_e$ (and uniquely determined quantum cuboid intertwiners) for boundary edges $e$, then the physical norm of such a state is given by
\begin{eqnarray}\label{Eq:PhysicalNorm}
\langle\eta(\phi)|\eta(\phi)\rangle_{\rm phys}\;\sim\;\sum_{j_f}\prod_v\widehat{\mathcal{A}}_v\prod_{f_e\in \partial\Gamma}\delta_{j_{f_e},j_e},
\end{eqnarray}

\noindent where $f_e$ is the unique face whose intersection with the boundary is $e$. Note that in (\ref{Eq:PhysicalNorm}) the boundary of the manifold has two connected components: the initial and the final spatial hypersurface. In an abuse of notation, $\phi$ is considered to be a state on both.

\begin{figure}[hbt!]
\begin{center}
\includegraphics[width=0.48\textwidth]{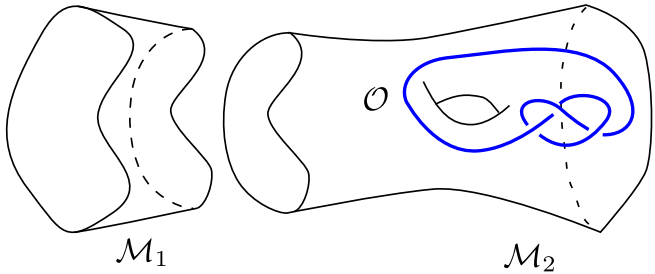}
\end{center}
\caption{The separation of space-time $\mathcal{M}\simeq\mathcal{M}_1\#\mathcal{M}_2$. \label{Fig:PIP_01}}
\end{figure}

This physical norm indicates how much the respective spin network contributes to the overall path integral, as we argue in the following.

Consider a manifold $\mathcal{M}$ with boundary $\Sigma_i\sqcup\Sigma_f$, being the product $\mathcal{M}=\mathcal{M}_1\#\mathcal{M}_2$ of two manifolds in the sense of cobordisms (see figure \ref{Fig:PIP_01}). If the Hilbert spaces are all finite-dimensional, then $Z$ provides a functor in the cobordism category, satisfying
\begin{eqnarray}\label{Eq:DecompositionOfSpaceTime}
Z(\mathcal{M})\;=\;Z(\mathcal{M}_1)\;Z(\mathcal{M}_2).
\end{eqnarray}

\noindent For any observable $\mathcal{O}$, its matrix element in boundary states $\Psi^{i, f}=\eta(\phi^{i,f})$ therefore satisfies
\begin{eqnarray}\nonumber
\langle \Psi^{i}|\mathcal{O}|\Psi^f\rangle=\sum_{\phi_n}\langle \Psi^{i}|\eta(\phi_n)\rangle_{\rm phys}\langle \eta(\phi_n)|\mathcal{O}|\Psi^f\rangle_{\rm phys}.\\[5pt]
\label{Eq:PhysicalNorm01}
\end{eqnarray}

\noindent Note that in this last equation, the sum is being performed over an orthonomal basis $\{\phi_n\}$ in the \emph{kinematical} Hilbert space $\mathcal{H}_{\rm kin}$. By using the Cauchy-Schwarz inequality, each term in that sum is dominated by a constant times $\|\eta(\phi_n)\|^2_{\rm phys}$. In other words, it can be expected that terms with a very small physical norm do contribute very little to the matrix element.

This argument can be generalized for arbitrary splitting of the manifold: The two-complex $\Gamma$ embedded in $\mathcal{M}$ is separated into two two-complexes $\Gamma_1\#\Gamma_2$. Then, each term in the state sum for $Z(\mathcal{M})$, i.e.~each collection of spins (and intertwiners) on $\Gamma$, can be found as part of a term in the sum (\ref{Eq:PhysicalNorm01}). In other words, in the general state sum for $Z(\mathcal{M})$, one can approximately neglect the sum over all states which, for some decomposition (\ref{Eq:DecompositionOfSpaceTime}) would result in a boundary spin network with small physical norm.

Therefore, it is of interest for the evaluation of the state sum, to identify boundary states with small physical norm. By excluding the corresponding states from the state sum, one can drastically simplify the state sum, making only small approximation errors. In particular, the error can be estimated by the physical norm of the excluded state.

It should be noted that the whole argument has to be slightly refined in the case of infinite-dimensional boundary Hilbert spaces. In that case, the state sum $Z(\mathcal{M})$ does not exist as an operator on $\mathcal{H}_{\rm kin}$, but at most as a quadratic form on the dense subset $\mathcal{D}_{\rm kin}$. In particular, an equation like (\ref{Eq:PhysicalNorm01}) will not hold in a straightforward sense. Still, since in this case the state sum will in many cases be defined as a limit of expressions which satisfy some sort of (\ref{Eq:PhysicalNorm01}), one can still expect the states with small (in particular those with vanishing) physical norm to be negligible.

\subsection{Torsion of boundary spin networks}

Motivated by this philosophy, we now compute the physical norm of boundary states for the special case of $\mathcal{M}\simeq [0, 1]\times T^3$. Using the geometric interpretation used in section \ref{Sec:Semiclassical}, we will show that one can define boundary states which are non-geometric in the sense that they contain non-vanishing torsion. Furthermore, we will show that boundary states with large torsion have small physical norm.

In the continuum, the torsion tensor $T$ is defined via $e_{a}^IT_{bc}^c=D_{[b}e_{c]}^I-[e_b, e_c]^I$. Geometrically, $T$ describes the failure of an infinitesimal square to close. After going around an infinitesimal parallelogram spanned by vectors $X^a$ and $Y^a$, one gets translated by the vector $T^a_{bc}X^bY^c$ (and rotated by the matrix $R^{a}{}_{bcd}X^cY^d$).
\begin{figure}[hbt!]
\begin{center}
\includegraphics[scale=0.6]{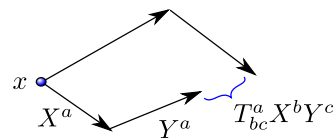}
\end{center}
\caption{Torsion measures the failure of an infinitesimal parallelogram to close.\label{Fig:Torsion01}}
\end{figure}

Taking literally the geometric interpretation of boundary geometries as provided by coherent polyhedra, this suggests a way to identify torsion in the boundary spin networks we use in our analysis.

The spatial hypersurfaces on the boundary are separated into cuboids, the areas of which are given in terms of the spins of the spin network function dual to it. This gives a geometric meaning to each cuboid, since its edge lengths are uniquely determined by its areas. These cuboids do not necessarily fit together without curving them, since for any square two of them are meeting, only the two respective areas agree, not necessarily the individual edge lengths.

\begin{figure}[hbt!]
\begin{center}
\includegraphics[width=0.48\textwidth]{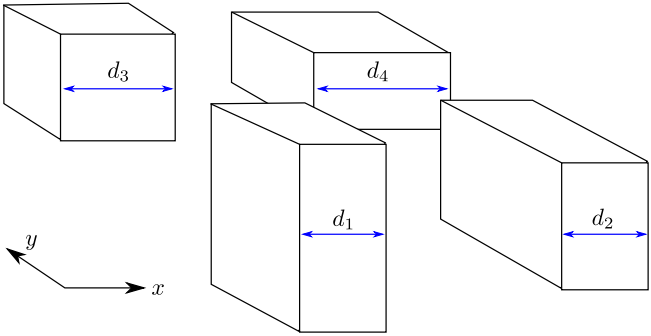}
\end{center}
\caption{Due to the non-geometric properties of twisted (spinning) geometries, the boundary geometry can contain torsion. In this case, $T_{xy}^x=\frac{1}{2}(d_1+d_2-d_3-d_4)$.\label{Fig:Torsion02}}
\end{figure}

Consider a path going along a minimal square in the boundary spin network function (say, in the $xy$-plane), i.e.~passing subsequently through the centres of four cuboids all meeting at one edge of the cubulation. The flat geometry suggests that each pair of neighboring sides are orthogonal to each other. So opposing sides are parallel, but do not necessarily have the same length. The difference between the two lengths (say, in $x$ direction) are precisely given by the $T_{xy}^x$-component of the torsion tensor (see figure \ref{Fig:Torsion02}).

\begin{figure}[hbt!]
\begin{center}
\includegraphics[width=0.48\textwidth]{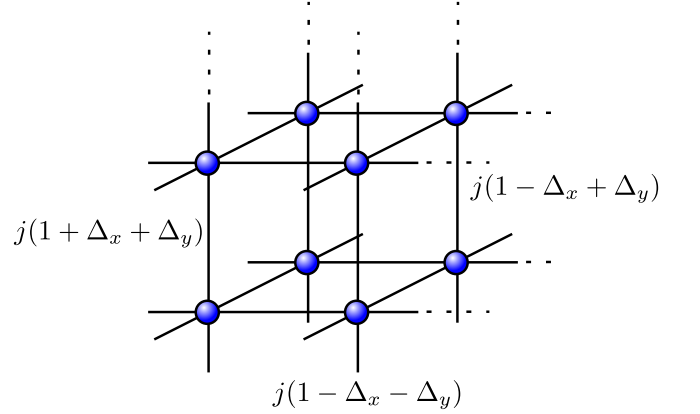}
\end{center}
\caption{Boundary spin network with torsion: the four vertical lines (dual to the $xy$-squares in the boundary cubulation) carry spins $j(1\pm\Delta_x\pm\Delta_y)$, while all horizontal lines (dual to the $xz$ and $yz$ squares) carry the spin $j$.\label{Fig:Torsion03}}
\end{figure}

With this in mind, we can construct spin networks which describe boundary geometry with torsion. For this, we choose a boundary of $2\times 2\times 2$ cuboids. The spin network under consideration will have one (large) $j$ distributed among the edges, apart from those on squares lying in the $xy$-plane. There are four of those, and they will each have one of the four possibilities of $j(1\pm\Delta_x\pm \Delta_y)$ (see figure \ref{Fig:Torsion03}). It can be readily seen that the components of the torsion tensor are $T_{xy}^x=2\Delta_x$, and $T_{xy}^y=2\Delta_y$.

\subsection{Results}

In our analysis, we have computed the physical norm (\ref{Eq:PhysicalNorm}) of the state $\psi_{\Delta_x,\Delta_y}$ by using a hypercubic lattice of $2\times 2\times 2\times N$, resulting in a norm
\begin{eqnarray}
\|\psi_{\Delta_x,\Delta_y}\|_{N, \alpha}^2\;=\;\sum_{\text{timelike} j_f}\prod_v\hat{\mathcal{A}}_v,
\end{eqnarray}

\noindent where the sum is being performed over all spins on time-like squares, while the spins over space-like squares are fixed by the lattice symmetry condition (\ref{Eq:LatticeSymmetry}). This means that the physical norm satisfies
\begin{eqnarray}
\|\psi_{\Delta_x,\Delta_y}\|^{2}_{N, \alpha}\;=\;\|\psi_{\Delta_x,\Delta_y}\|^{2N}_{1, \alpha},
\end{eqnarray}

\noindent so we can restrict ourselves to $N=1$. In this case, there are six free spins in total, over which we integrate numerically, using Mathematica 10. We show the results, depending on the torsion of the boundary state, which is parametrised by $\Delta_x,\Delta_y$, for three different values of $\alpha$. The parameters $\Delta_x,\Delta_y$ are varied from $-0.4$ to $0.4$ in steps of $0.1$, respectively. The results are normalized so that the state with $\Delta_x=\Delta_y=0$ has physical norm one.

The numerical calculation of the physical norm shows very similar features for all three values of $\alpha$: The state with $\Delta_x=\Delta_y=0$ have maximal physical norm (normalized to 1), while states with higher torsion have smaller physical norm. The norm of a kinematical state $\psi_T$ with torsion $T$ is suppressed with a factor
\begin{eqnarray}
\|\psi_T\|_{\rm phys}^2\;\sim\; e^{-C_\alpha |T|^2}.
\end{eqnarray}

\begin{widetext}

\begin{figure}[hbt]
\includegraphics[width=\textwidth]{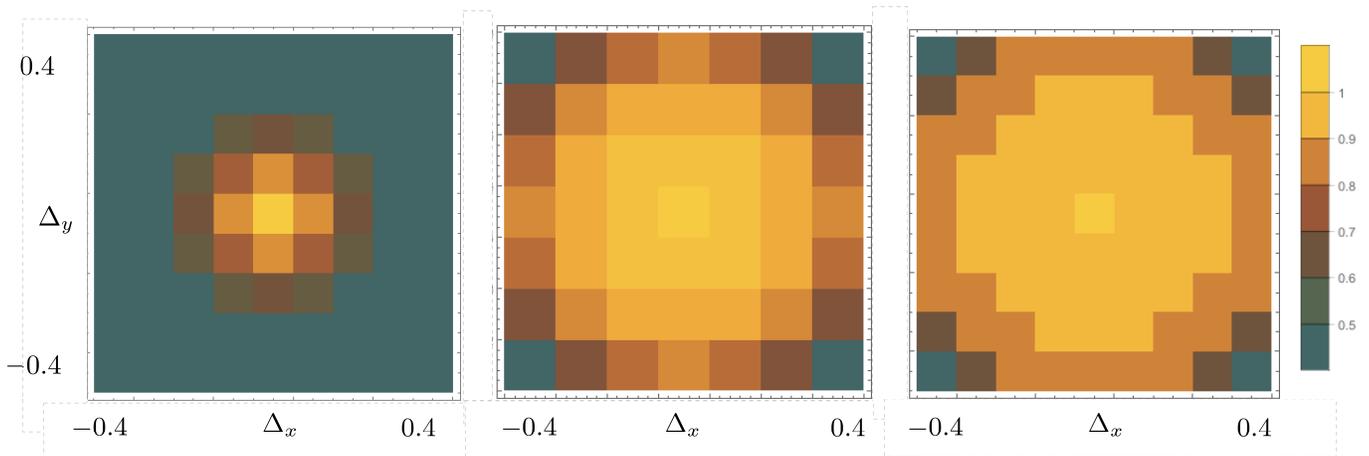}
\caption{Physical norm of states, depending on their torsion (color-coded). The plots are given for (from left to right) $\alpha=4$, $\alpha=0.8$, and $\alpha=0.6$.\label{Fig:TorsionAlpha}}
\end{figure}

\end{widetext}

\begin{figure}[hbt!]
\begin{center}
\includegraphics[width=0.48\textwidth]{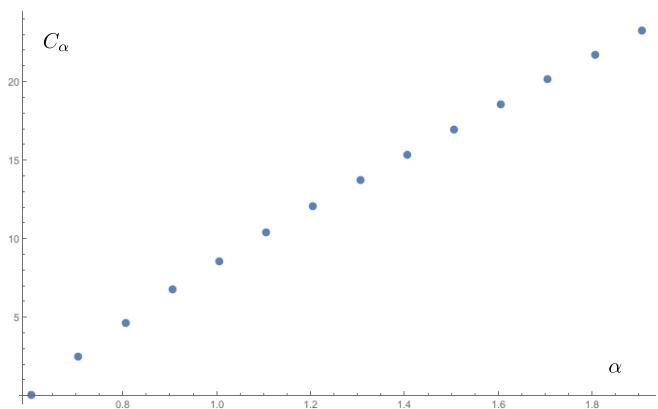}
\end{center}
\caption{The inverse widths of the Gaussians in figure \ref{Fig:TorsionAlpha}, depending on the coupling constant $\alpha$. Numerical investigations show that $C_\alpha = 17.22\alpha-10.45 $ is an excellent fit.\label{Fig:TorsionFinal}}
\end{figure}

\noindent The rate of suppression $C_\alpha$ depends on $\alpha$, and the connection between the two is depicted in figure \ref{Fig:TorsionFinal}. As one can see, the larger $\alpha$, the more one can disregard torsion.

It should be noted that our definition of torsion is a non-local property, in the sense that it requires more than one vertex to define it. It is closely related to the nongeometricity $\xi$, in the sense that torsion can only exist if there is non-geometricity in the vertices. However, the converse is not true: One can show that there are semiclassical configurations which are highly non-geometric, but have vanishing torsion. In that sense torsion-freedom is a weaker condition than geometricity.

\section{Summary, Conclusion, and Outlook}

In this article, we have investigated properties of the quantum gravity path integral as given by the EPRL-FK spin foam model. We used the method of symmetry restriction, which resulted in considering the sum only over spins and intertwiners which satisfy certain symmetry requirements.

The discretization of space-time we considered consisted of a 4-dimensional hypercubic lattice, embedded in $\mathcal{M}\simeq [0,1]\times\mathbbm{R}^3$. The intertwiners were restricted to the set of \emph{quantum cuboids}, i.e. coherent polyhedra with six faces, where the pairs of normals were negative of each others.

Due to this choice, every edge was six-valent, and opposite faces had to have equal spins. This enforced a large amount of symmetry throughout the lattice, restricting the configurations immensely. Despite the drastic simplification, we gained several insights into the path integral, which should be useful in future analyses.

The Barbero-Immirzi parameter $\gamma$ proved to be irrelevant for the semiclassical analysis, which was a direct consequence of the fact that only flat configurations were considered. For the same reason, Newton's constant could be set to $1$, and did not appear in the analysis.

The only relevant coupling constant in our analysis was $\alpha$, the power of the face amplitude. Several properties of the path integral crucially depend on it.

\subsection*{Non-geometricity is suppressed in the path integral}

The states in spin foam models have a certain description in terms of twisted (or spinning) geometries. It is well-known that these do not necessarily correspond to four-dimensional metrics. We identified the constraints that states had to satisfy to describe a four-dimensional metric geometry, in terms of local constraints on the spins (i.e.~areas). The constraints were equivalent to the existence of an unambiguously defined $4$-volume of the vertex.

Deviation from this constraints were identified as non-geometricity $\xi$. Its physical role in the path integral crucially depends on the value of the parameter $\alpha$. For $\alpha<\alpha_\xi\approx 0.49$, out of all hypercuboids of a fixed $4$-volume, the non-geometric ones contributed the most. On the other hand, for $\alpha>\alpha_\xi$, the geometric configurations dominate the non-geometric ones.  In particular, for large enough values of $\alpha$, non-geometricity is duely suppressed (see also the analysis in \cite{Livine:2006it}).

Furthermore, the effective action for the state sum in the large $j$ limit contains a mass term $m_\xi^2$ for the non-geometricity $\xi$, which grows linearly with $\alpha$.

All of this analysis results from investigations around a regular hypercuboid -- for irregular hypercuboids, as far as we can see, the qualitative features of the analysis remain the same, while e.g.~the numerical value of $\alpha_\xi$ changes slightly. As a general rule, it seems that non-geometricity is suppressed in the path integral, the stronger the larger $\alpha$ is.

\subsection*{Diffeomorphism symmetry in the path integral}

As it turned out, despite the simplification of the path integral due to the restriction to quantum cuboids, the remaining state space is large enough so that an Abelian subgroup of the diffeomorphisms acts on it. In the geometric large $j$-limit, this action can be interpreted as vertex-translation symmetry. It can be defined even on non-geometric configurations. Some of them, called semi-geometric, were such that the diffeomorphism action could be defined in a volume preserving way. The states could therefore be separated into three categories
\begin{eqnarray*}
\begin{array}{c}\text{all}\\[-5pt]\text{states}\end{array}\;\subset\;
\begin{array}{c}\text{semi-}\\[-5pt]\text{geometric}\end{array}\;\subset\;\text{geometric.}
\end{eqnarray*}

\noindent As it turned out, the value of the coupling constant $\alpha$ crucially influences the behaviour of the amplitude under the action of the diffeomorphisms. There is a critical value $\alpha_c\approx 0.61$, which separates two regions in phase space. The region $\alpha<\alpha_c$ is such that those configurations dominate the path integral, in which almost all of the $4$-volume is concentrated in one time-slice, while the other time-slices does contain almost no $4$-volume at all. On the other hand, for $\alpha>\alpha_c$, the state with the major contribution to the path integral is the one in which the $4$-volume is separated equally among all vertices.

A state (or a collection of states) which dominates the state sum can be interpreted as ``vacuum'', in accordance with lattice gauge theory nomenclature. The reason is that it corresponds to ground states of the Hamiltonian, and the state sum describes fluctuations around it. We adopt this notion in the following discussion, and call the state (or the collection of them) which dominates the path integral as vacuum.

The vacuum state in the two phases is quite different: while below the critical point $\alpha=\alpha_c$ it is a superposition of space-times with one ``fat time-slice'' and many ``thin time-slices'', above the critical point the vacuum is described by a regular lattice, i.e.~where all spins throughout the hypercubic lattice are equal.

These two different states can be viewed as two different realizations of flat Euclidean space (i.e.~a flat geometry on a torus). In one the vacuum describes the hypercubic lattice as very regularly embedded in the torus, while the other is a superposition of many different irregular ways in which the hypercubic lattice is embedded. Equivalently, the $\alpha>\alpha_c$ case can be seen as the completely regular metric $ds^2=\sum_i d\phi_i^2$ (where $\phi_1,\ldots\phi_4$ are four angular coordinates on the 4-torus), while the highly irregular configurations dominating in $\alpha<\alpha_c$ can be regarded as metrics of the form $ds^2=\sum_i f_i(\phi_i)d\phi_i^2$ for some highly flutuating functions $f_1(\phi_1),\ldots, f_4(\phi_4)$. Obviously, these two types of metrics are diffeomorphism equivalent.

So the point $\alpha=\alpha_c$ separates two  different vacua, both of which break diffeomorphism symmetry (i.e.~the state-sum picks out different representatives of the diffeomorphism orbit).
%

There is furthermore some reason to assume that the critical point $\alpha=\alpha_c$ could constitute a phase transition of second order. The reason is that, at this point, vertex translation symmetry is nearly realized: to a certain numerical accuracy, all diffeomorhism-equivalent metrics describing the same geometry contribute the same value to the path integral. If one interprets vertex translation symmetry as coming from diffeomorphisms, this would indicate that here one could find a diffeomorphism-invariant fixed point, even in the full path integral. Since diffeomorphism-invariance means that correlation lengths diverge (in terms of lattice distance), this could also indicate the existence of a continuum limit at this critical point. Obviously, this deserves a lot more investigation, and is an exciting prospect for the search of the theory of quantum gravity.

\subsection*{Torsion and the physical inner product}

In the last part of the article, we investigated the physical inner product as defined by the symmetry-restricted state sum. In particular, we looked at the dependence of the physical norm of kinematical states, depending on the amount of torsion that was contained in its boundary geometry. We defined the torsion in terms of the naive geometric interpretation of the coherent polyhedra as cuboids in the large $j$-limit. It should be noted that, at this point, the connection to the quantum continuum torsion usually discussed in the literature \cite{Dittrich:2008ar, Dittrich:2012rj, Haggard:2012pm, Anza:2014tea} is not completely understood at this point. This certainly warrants further study.

What we found was that kinematical boundary states with high torsion have suppressed physical norm. The rate of suppression grows with increasing $\alpha$. Since torsion, in the way that we have defined it, on the boundary implies non-geometricity in the bulk, this fits together with the observation that, for higher values of $\alpha$, non-geometricity is suppressed in the path integral. However, the statement is stronger than that. The physical norm is not just the value of the amplitude for a specific state, but includes a sum over states.

This has an important implication: To a good approximation, one can restrict the state sum to only those states, which are such that if the $2$-complex is divided along any $3d$ hypersurface, the determined boundary geometry on that hypersurface has no torsion. This is a non-local condition on states, given the naive definition of torsion we have used, in the sense that it is not a restriction of spins and intertwiners or vertices, but a property of the labels of bunches of neighboring vertices. Understanding this condition better, to find an effective approximation and restriction of the state sum, would be very interesting. It could also allow to get a much better and efficient way of computing state sums, at least for not too small values of $\alpha$.

\begin{widetext}
\begin{center}
\begin{figure}[h]
\includegraphics[width=0.75\textwidth]{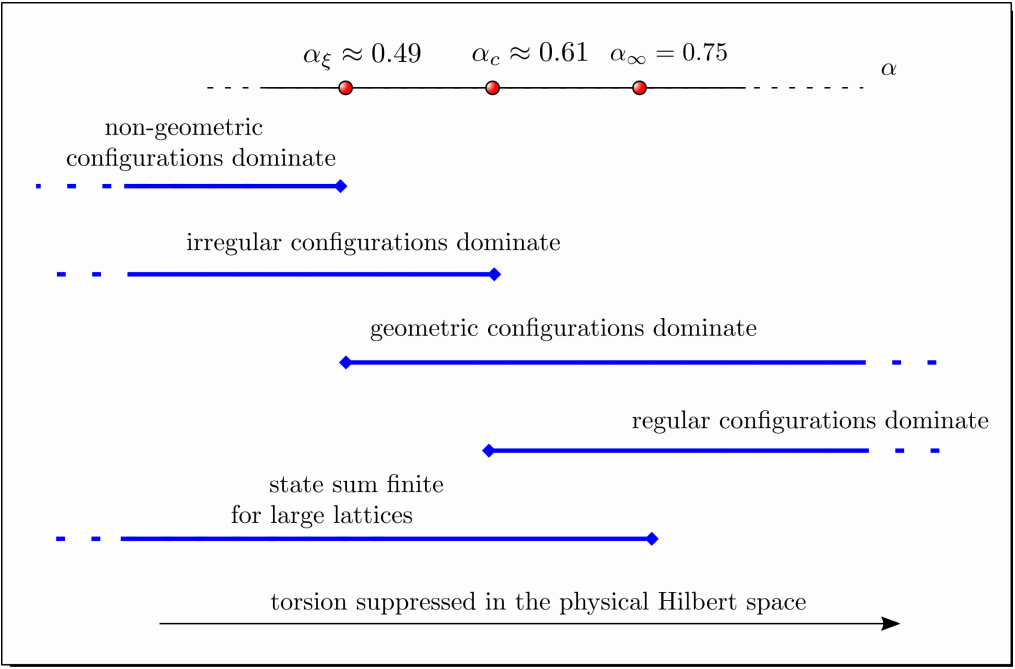}
\caption{Different phases of the EPRL-FK model, depending on the coupling constant $\alpha$. It should be noted that the precise numerical values should not be taken too literally, due to the simplifications and choices we have made in our analysis. However, we do expect that a qualitatively similar phase diagram could be found also in the full path integral.\label{Fig:Conclusion}}
\end{figure}
\end{center}
\end{widetext}

\subsection*{Outlook}

Although some interesting statements about the spin foam state sum could be derived in this article, it remains to be shown that these statements survive if more states are included than just the ones containing quantum cuboids. In particular, it is paramount to include states which allow for the summation over curvature degrees of freedom, to arrive at a more realistic approximation to the whole state sum.

Furthermore, it should be noted that we used the unrenormalized EPRL-FK amplitude. In working on a fixed lattice, this disregards radiative corrections from finer lattices. \footnote{In the GFT picture, this would correspond to sum over more than one triangulation.} What one should actually do is use the renormalized amplitude instead, which serves as an effective action on the fixed lattice. This should in particular contain more coupling constants, such as e.g.~the cosmological constant or prefactors in front of higher curvature and even non-local terms.

At this moment, however, this amplitude does not exist. In fact, the situation is such that very little is known about the renormalization group flow of the EPRL-FK model. Performing actual calculations in this direction with the full state sum might prove very challenging, both analytically and numerically. This is why, to tackle this problem, the methods shown in this article, could prove useful to that end.


\begin{acknowledgments}
This work was funded by the project BA 4966/1-1 of the German Research Foundation (DFG). The authors are indebted to Sebastian Kl\"oser and Giovanni Rabuffo, for pointing out a computational error.
\end{acknowledgments}

\bibliography{bibliography}

\end{document}